\documentclass[journal]{IEEEtran}
\ifCLASSINFOpdf
\else
\fi
\usepackage{import}

\usepackage{float}

\usepackage{algorithm}
\usepackage{algpseudocode}

\usepackage{amsthm}
\newtheorem{mytheo}{Theorem}
\newtheorem{exmp}{Example}

\usepackage{graphicx} 
\usepackage{epsfig} 
\usepackage{times} 
\usepackage{amsmath} 
\usepackage{amssymb}  
\usepackage{amsthm}
 \usepackage{pgfplots}
\usepackage{xcolor} 

\usepackage{caption}
\usepackage{subcaption}

\usepackage{multirow}

\usepackage{tikz}
\usetikzlibrary{positioning, arrows, calc, shadings}

\tikzstyle{blockdiag}	= [node distance=5mm, >=stealth', semithick]
\tikzstyle{block}			= [draw, rectangle, minimum width=1cm, minimum 
height=.8cm]
\tikzstyle{sum} = [draw,circle,inner sep=0pt, minimum size=6pt]
\tikzstyle{connector} = [draw,circle,inner sep=0pt, minimum size=2pt, 
fill=black]

\usepackage{array}
\newcommand{\norm}[1]{\left\|#1\right\|}
\newcommand{\abs}[1]{\left|#1\right|}
\newcommand{\field}[1]{\mathbb{#1}}
\newcommand{\R}{\field{R}}
\newcommand{\RH}{\field{RH}_\infty}
   
\newcommand{\RL}{\field{RL}_\infty}
\newcommand{\C}{\field{C}}

\newcommand{\bmtx}{\begin{bmatrix}}
\newcommand{\emtx}{\end{bmatrix}}
\newcommand{\bsmtx}{\left[ \begin{smallmatrix}} 
\newcommand{\esmtx}{\end{smallmatrix} \right]} 
\newcommand{\bmatarray}[1]{\left[\begin{array}{#1}}
\newcommand{\ematarray}{\end{array}\right]}

\definecolor{blue1}{RGB}{222,235,247}
\definecolor{blue2}{RGB}{158,202,225}
\definecolor{blue3}{RGB}{49,130,189}

\definecolor{mycolor1}{rgb}{0.00000,0.44700,0.74100}%
\definecolor{mycolor2}{rgb}{0.85000,0.32500,0.09800}%
\definecolor{mycolor3}{rgb}{0.92900,0.69400,0.12500}%
\definecolor{cubicpurple}{rgb}{0.51510,0.04820,0.66970}%
\definecolor{cubicblue}{rgb}{0.33370,0.56500,0.90560}%
\definecolor{cubicgreen}{rgb}{0.31980,0.84920,0.39560}%
\definecolor{cubicyellow}{rgb}{0.80000,0.92550,0.35290}%
\definecolor{cubicred}{rgb}{0.97650,0.58870,0.35690}%
 \definecolor{lgray1}{rgb}{0.3,0.3,0.3}%
 \definecolor{lgray2}{rgb}{0.6,0.6,0.6}%
 \definecolor{lgray3}{rgb}{0.80000,0.8,0.8}%
\begin{document}
%
\title{Finite Horizon Worst-Case Analysis of Linear Time-Varying Systems Applied to Launch Vehicle}
%
%
%


\author{Felix Biert\"umpfel, 
           Nantiwat Pholdee,
          Samir Bennani,
       Harald Pfifer~
\thanks{Manuscript received March 2, 2021; revised xxx.}%
\thanks{This work is partially funded by ESA through the Networking/Partnering Initiative contract No. 4000123233 and The National Research Council of Thailand (NRCT5-RSA63003-06)}%
\thanks{F. Biert\"umpfel and H. Pfifer are with the Technische Univerist\"at Dresden, Institute of Aerospace Systems, Chair of Flight Mechanics and Control, 01307, Dresden (felix.biertuempfel@tu-dresden.de; harald.pfifer@tu-dresden.de)}%
\thanks{N. Pholdee is with the Khon Kaen University, Sustainable and Infrastructure Research and Development Center, 40002, Khon Kaen, Thailand (nantiwat@kku.ac.th)}%
\thanks{S. Bennani is with the European Space Agency, ESTEC, ESA, 2201AZ, Noordwijk, Netherlands (samir.bennani@esa.int)}}%

%
%

\markboth{Journal of \LaTeX\ Class Files,~Vol.~14, No.~8, August~2015}%
{Shell \MakeLowercase{\textit{et al.}}: Bare Demo of IEEEtran.cls for IEEE Journals}
%



\maketitle

\begin{abstract}
This paper presents an approach to compute the worst-case gain of the interconnection of a finite time horizon linear time-variant system and a perturbation. The input/output behavior of the uncertainty is described by integral quadratic constraints (IQCs). A condition for the worst-case gain of such an interconnection can be formulated using dissipation theory as a parameterized Riccati differential equation, which depends on the chosen IQC multiplier. A nonlinear optimization problem is formulated to minimize the upper bound of the worst-case gain over a set of admissible IQC multipliers. This problem can be efficiently solved with a custom-tailored logarithm scaled, adaptive differential evolution algorithm. It provides a fast alternative to similar approaches based on solving semidefinite programs. The algorithm is applied to the worst-case aerodynamic load analysis for an expendable launch vehicle (ELV). The worst-case load of the uncertain ELV is calculated under wind turbulence during the atmospheric ascend and compared to results from nonlinear simulation.
\end{abstract}


%
\IEEEpeerreviewmaketitle

\section{Introduction}
\label{sec:intro}
\IEEEPARstart{N}{umerous} systems follow a predefined trajectory with a given start and terminal point during their nominal operation. Examples are various and include robot arms \cite{Hosovsky2016}, aircraft on final approach, as well as space applications such as launch vehicles \cite{Haeussermann1970}, or vehicles for atmospheric re-entry \cite{Juliana2004}.
The dynamics of these examples can all be represented as time-varying systems. Most of these systems can be described by a linearization along a given operating trajectory. Consequently, they are linear time-varying (LTV) systems, i.e. linear systems whose state matrices depend solely on time. 
Examples for systems representable by the finite time horizon form are, besides the opening examples, terminal guidance systems \cite{HeWa15} and controlled swarm robots \cite{MelVa12}. Finite horizon LTV systems are formally introduced in Section \ref{ss:LTVsys}. 
Closely related to these types of systems is another class of linear time-varying systems, namely periodic ones. Systems coverable by the periodic form
are, for example, the flapping of a helicopter rotor blade in forward flight \cite{Wer91}, wind turbines \cite{Ossmann2017},
and spinning satellites \cite{BalYu97}. 

In the literature, numerous approaches to calculate the robustness of
uncertain periodic LTV systems are given \cite{Dul1, Kim1, Ossmann2019}.
In contrast, the analysis of finite horizon LTV systems is significantly less exploited.
The work in \cite{Joe1} proposes robustness measures for finite time trajectories using
integral quadratic constraints (IQCs) to represent the uncertainties. The presented approach does not consider disturbances.
The analysis via gap metrics is covered in \cite{Can3} or \cite{CanPfi1}.

In this paper, a worst-case analysis for uncertain LTV is proposed based on the extension of a finite horizon formulation of the bounded real lemma (BRL) given in \cite{Green1995}. The strict BRL proposed provides an analysis condition for nominal LTV systems under external disturbance. It is based on the solvability of a Riccati differential equation (RDE), which can be turned into a computational feasible problem. 
The system's perturbations are represented by integral quadratic constraints (IQCs). First introduced by \cite{MegRa1}, IQCs
presents a general framework for robustness analyses. They are able to cover numerous types of perturbations such as uncertainties, hard nonlinearities, or infinite-dimensional systems. In \cite{MegRa1}, a broad range of multipliers defined in the frequency
domain is given. Due to the frequency domain interpretation in \cite{MegRa1}, only nominal systems that are linear time-invariant can be covered. 
More recently, the IQC framework has been studied in the time domain such as \cite{Seiler2015}, \cite{Can2}, and \cite{Veenman2016}. This time-domain formulation of IQCs is given in Section \ref{ss:IQC}. 
The time-domain approach allowed the extension to cover linear parameter
varying \cite{PfiSei2}, \cite{Scher3} or nonlinear polynomial systems \cite{Pfi1} under perturbations. They provide sufficient conditions on worst case input/output gains based on a BRL-like condition. 
These advances opened the IQC framework for LTV systems in \cite{Moore1}. The analysis conditions are based on the dissipation inequality conditions presented in \cite{PfiSei2} for uncertain LPV systems. This LTV robustness analysis framework is described in Section \ref{ss:LTVRob}.
In \cite{Moore1}, the analysis condition is only enforced on a finite-dimensional grid to overcome the problem's infinite nature. In \cite{Seiler2019}, the linear matrix inequality (LMI)-based approach in \cite{Moore1} is extended using an equivalent RDE formulation of the LMI conditions. The LMI and RDE are solved iteratively to mitigate the effect of the gridding and calculate a less conservative upper bound on the worst-case gain. This approach was successfully demonstrated for academic examples over short time horizons.

On the contrary, in this paper, the focus is on complex industrial examples analyzed over long time horizons for which no examples in the literature exist. The proposed approach exclusively utilizes the RDE condition. It exploits the direct dependence of the RDE on the IQC multiplier/parameterization. This leads to a nonlinear optimization problem centered around solving a parameterized RDE and optimizing over the IQC multiplier/ parameterization. It is efficiently solved using a tailored algorithm based on logarithmically scaled adaptive differential evolution with linear population size reduction (Log-L-SHADE). The population size scheme is based on the original L-SHADE \cite{Tanabe2014}, but the logarithmic scale significantly reduces the search space. Furthermore, characteristics of the optimization problem and structure of the applied IQCs are exploited to reduce the computational effort further. A detailed description of the novel nonlinear program is given in \ref{sec:CompAppr}.

The algorithm's applicability to industry-relevant examples is demonstrated in a detailed analysis presented in Section \ref{sec:Example}. There, the worst-case aerodynamic load on an expendable launch vehicle (ELV) during its ascent under atmospheric turbulence is calculated. The analysis utilizes an advanced turbulence disturbance model adjusted to the finite horizon LTV framework's requirements imposed by the BRL. Furthermore, an adequate uncertainty set in the IQC framework is introduced. To finish, a Monte Carlo simulation on the corresponding nonlinear simulation over the allowable uncertainty set and turbulence profiles  is conducted. It is used to validate that the finite horizon LTV analysis using IQCs provides a valid and not too conservative upper bound for the nonlinear model in a fraction of time.

Thus, this paper contributes an efficient algorithm to calculate the worst-case gain of uncertain finite horizon LTV systems. It is based on the solvability of a parameterized RDE and nonlinear optimization with a specialized solver that exploits the problem's structure.
The approach's feasibility is demonstrated on a high fidelity example in the form of a worst-case loads analysis of a space launcher during atmospheric ascent.


\section{Notation}
\label{sec:Notation}
In the course of this paper, $\R$ and $\C$ denote the set of real and complex numbers respectively.
The set of rational functions with real coefficients that are proper and have no poles on the imaginary axis is
denoted by $\RL$. Herein, $\RH$ is the subset of functions in $\RL$ that are analytical in the closed right half of the complex plane.
The sets of $m \times n$ matrices whose elements are in $\C$, $\RL$ and $\RH$ are denoted $\C^{m \times n}$, $\RL^{m \times n}$ and $\RH^{m \times n}$ respectively. Vectors are described by a single superscript, e.g., $\R^n$ being the set of vectors whose
elements are in $\R$. A vertical concatenation of the vectors $x \in \R^n$ and $y \in \R^m$ is denoted by $[x;y] \in \R^{n+m}$.
The set of $n \times n$ symmetrical matrices is denoted by $\mathbb{S}^n$. $\R^+_0$ denotes the set of positive real numbers including
zero. The set of positive real numbers excluding zero is denoted by $\R^+$. For $z \in \C$, $\bar{z}$ denotes the complex conjugate of $z$. The transpose of a matrix $M \in \C^{m \times n}$ is denoted by $M^T$. 
Furthermore, structured matrices $C$ are expressed concisely via the Kronecker product $C = A \otimes B = (a_{ij} \cdot B)$. 
The size of signals in this paper is described by the Lebesgue 2-norm \cite{Green1995}.
Let $ S :=\{ v: \R \rightarrow \R^{n\times n}\}$ denote the set of (Lebesgue) measurable signals, with the
subspace
\begin{equation}
\label{eq:Leb1}
S_+ :=\{ v \in S \,| \, v(t) = 0, \forall \, t < 0\}
\end{equation}
The finite horizon Lebesgue 2-norm can be defined as:
\begin{equation}
\label{eq:Leb2}
\norm{v}_{2[0,T]} = \bigg[ \int_0^T v^T(t)v(t)\,dt \bigg]^\frac{1}{2}
\end{equation}
Accordingly, the Lebesgue 2-space $L_2[0,T]$ representing signals on $[0, T]$ with finite energy is defined as
\begin{equation}
\label{eq:Leb3}
L_2[0,T] : = \{ v \in S_+\, | \, \norm{v}_{2[0,T]} < \infty \}
\end{equation}
As any continuous signal on $[0, T]$ is bounded, it therefore lies in $L_2[0, T]$.

\section{Background on the Robustness Analysis of Finite Horizon LTV Systems}
\label{sec:RobAnalysis}
\subsection{Finite Horizon Linear Time Varying Systems}\label{ss:LTVsys}
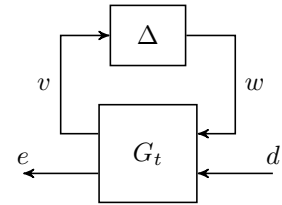
\begin{figure}[ht!]
\centering
\begin{tikzpicture}[blockdiag]
	
	\node[block,minimum width=1.3cm, minimum height=1.3cm](Plant) {$ G_t$};
	\node[block, above= of Plant] (Delta) {$\Delta$};

	\draw[<-] ($(Plant.south east)!.3!(Plant.north east)$) -- +(+1.0cm, 0) 
	node[above, name=e]{$d$};
	\draw[->] ($(Plant.south west)!.3!(Plant.north west)$) -- +(-1.0cm, 0) 
	node[above, name=d]{$e$};
	
	\draw[<-] ($(Plant.south east)!.7!(Plant.north east)$) -- +(+0.5cm, 0) |- 
	node[near start, right]{$w$} (Delta);	
	\draw[->] ($(Plant.south west)!.7!(Plant.north west)$) -- +(-0.5cm, 0) |- 
	node[near start, left ]{$v$} (Delta);
	
\end{tikzpicture}
\caption{Interconnection of LTV system and perturbation}
\label{fig:UncSys}
\vspace{0pt}
\end{figure}

An uncertain system is defined by the feedback interconnection of a nominal LTV system $G_t$ and the perturbation $\Delta$ as shown in Fig. \ref{fig:UncSys}.
This interconnection is denoted by $F_u(G_t, \Delta)$, with the LTV system $G_t$ defined as
\begin{equation}
\label{eq:LTVsys}
\begin{split}
\dot{x}_{G_t}(t) &= A_{G_t}(t)x_{G_t}(t) + B_{G_t}(t) \bsmtx w(t) \\ d(t) \esmtx \\
\bsmtx v(t) \\ e(t) \esmtx &= C_{G_t}(t)x_{G_t}(t) + B_{G_t}(t) \bsmtx w(t) \\ d(t) \esmtx.
\end{split}
\end{equation}
In (\ref{eq:LTVsys}), $x_{G_t}(t) \in \R^{n_{x_{G_t}}}$, $d(t) \in \R^{n_d}$, and $e(t) \in \R^{n_e}$ denote the state, input, and output vector, respectively. The matrices $A_{G_t}$, $B_{G_t}$, $C_{G_t}$ and, $D_{G_t}$ are piecewise continuous locally bounded matrix-valued functions of time with the appropriate dimensions. To shorten the notation, the explicit time dependence is generally omitted in this paper. Therefore, LTI systems will be introduced as such. 
The uncertainty $\Delta: L_2^{n_v}[0,T] \rightarrow L_2^{n_w}[0,T]$ is a bounded and causal operator. 
$\Delta$ can describe hard nonlinearities like saturations, infinite dimensional operators like time delays, as well as dynamic and real parametric uncertainties.


\subsection{Integral Quadratic Constraints}\label{ss:IQC}
The input/output behavior of the perturbations is bounded using IQCs. IQCs were first introduced in the context of robustness analysis in the frequency-domain by \cite{MegRa1}.
More recently, the IQC framework has been investigated in the time-domain \cite{Veenman2016, Seiler2015, PfiSei2}. Time domain IQCs are generally distinguished into hard IQCs, which hold over all finite time horizons, and soft IQCs, which only hold over infinite horizons. Hard factorizations are used in e.g., \cite{PfiSei2} to assess the stability of LPV systems. The LTV analysis presented  here only requires the IQC to hold over the respective finite analysis horizon $[0,T]$. 

A time-domain IQC is defined by the filter $\Psi \in \RH^{n_z \times (n_v + n_w)}$ and a symmetric
matrix $M \in \mathbb{S}^{n_z}$. 
The short notation $\Delta \in IQC(\Psi, M)$ is used if the perturbation $\Delta$ satisfies
the IQC defined by $\Psi$ and $M$ over the interval $[0,T]$. 
This is the case if the output $z$ of the filter $\Psi$ fulfills the quadratic time constraint
\begin{equation}
    \label{eq:iqctd1}
    \int_0^T z(t)^T M z(t) \, dt \ge 0
\end{equation}
for all $v \in L_2[0,T]$ and $w = \Delta(v)$ given a finite interval $[0, T]$.
The IQC framework allows to include $k$ different perturbations $\Delta_i \in IQC(\Psi_i, M_i)$ in a single IQC 
by diagonally combining them.

\subsection{LTV Robustness Analysis}\label{ss:LTVRob}
\begin{figure}[h!]
\centering
\begin{tikzpicture}[blockdiag]
	
	\node[block,minimum width=1.3cm, minimum height=1.3cm](Plant) {$ G_t$};
	\node[block,dashed, above= of Plant] (Delta) {$\Delta$};
	\node[block, above right=of Delta, xshift=2mm] (Psi) {$\Psi$};

	\draw[<-] ($(Plant.south east)!.3!(Plant.north east)$) -- +(+1.0cm, 0) 
	node[above, name=e]{$d$};
	\draw[->] ($(Plant.south west)!.3!(Plant.north west)$) -- +(-1.0cm, 0) 
	node[above, name=d]{$e$};
	
	\draw[<-] ($(Plant.south east)!.7!(Plant.north east)$) -- +(+0.5cm, 0) |- 
	node[near start, right]{$w$} (Delta);	
	\draw[->] ($(Plant.south west)!.7!(Plant.north west)$) -- +(-0.5cm, 0) |- 
	node[near start, left ]{$v$} (Delta);
	
	\draw[->] (Delta.west) -- +(-0.3cm, 0) |- ($(Psi.south west)!.7!(Psi.north 
	west)$);
	\draw[->] (Delta.east) -- +(+0.2cm, 0) |- ($(Psi.south west)!.3!(Psi.north 
	west)$);
	
	\draw[->] (Psi.east)   -- +(.5cm,0)    node[above, name=z]{$z$};
	
\end{tikzpicture}
\caption{Feedback Interconnection LTV system $G_t$ and uncertainty $\Delta$}
\label{fig:fbic_LTV_IC}
\end{figure}
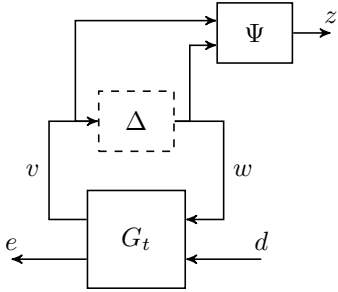

The feedback interconnection of an LTV system $G_t$ and 
uncertainty block $\Delta$ in the IQC framework is pictured in Fig. \ref{fig:fbic_LTV_IC}
The input $v$ and output $w = \Delta(v)$ of $\Delta$ are connected to the IQC filter
$\Psi$. Consequently, $\Delta$ is excluded from the interconnection as emphasized in Fig. \ref{fig:fbic_LTV_IC}.
Accordingly, $w$ is now treated as an external signal in the extended state space system $G$
\begin{equation}
  \label{eq:Pext}
  \begin{split}
   \dot{x}(t) &= A(t) \, x(t) + \bsmtx B_1(t) & B_2(t)\esmtx \bsmtx w(t) \\ d(t) \esmtx \\  
   \bsmtx z(t) \\ e(t)\esmtx &= \bsmtx C_1(t)\\ C_2(t) \esmtx x(t) + \bsmtx D_{11}(t) & D_{12}(t) \\ D_{21}(t) & D_{22}(t) \esmtx \bsmtx w(t) \\ d(t) \esmtx , 
  \end{split}
\end{equation}
where  $x(t) = [x_{G_t}(t)^T, x_{\Psi}(t)^T]^T\in \R^{n_{x_{G_t}}+n_{x_\Psi}}$ represents the state vector containing the states of $G_t$ and $\Psi$, $d(t) \in \R^{n_d}$ the external disturbance input vector, and $e(t) \in \R^{n_e}$ the performance output vector. 
The internal signal $v$ and external signal $w$ are subject to the time-domain constraint (\ref{eq:iqctd1}) enforced on the output of the IQC filter $z$.
Hence, the explicit formulation of $w = \Delta(v)$ is replaced by the time-domain inequality (\ref{eq:iqctd1}).

The robust performance of an uncertain LTV system in the IQC framework can then be quantified by worst-case finite horizon input/output gains. Specifically, two metrics are used in this paper. Firstly, the finite horizon worst case $L_2[0,T]$ to $\norm{e(T)}_{2}$ gain defined as follows:
\begin{equation}
\label{eq:E2PWC}
\begin{gathered}
  \| F_u(G_t,\Delta) \|_{2} := \sup_{ \Delta \in \textrm{IQC}(\Psi ,M)} \sup_{\substack{d \in
      L_2[0,T]\\ d \neq 0, x(0) = 0}} \frac{ \norm{e(T)}_{2}}{\norm{d(t)}_{2[0,T]}}.
\end{gathered}
\end{equation}
Geometrically interpreted, it describes the ball upper bounding the worst case output $e(T)$ over all $\Delta \in IQC(\Psi, M)$ for $\norm{d(t)}_{2[0,T]} = 1$ and the considered finite time horizon $[0, T]$ with $T \in [0, \infty)$.
The second performance measure is the finite horizon worst case $L_2[0,T]$ gain
\begin{equation}
\label{eq:L2WC}
\begin{gathered}
  \| F_u(G_t,\Delta) \|_{2[0, T]} := \sup_{ \Delta \in \textrm{IQC}(\Psi ,M)} \sup_{\substack{d \in
      L_2[0,T]\\ d \neq 0, x(0) = 0}} \frac{ \norm{e(t)}_{2[0,T]}}{\norm{d(t)}_{2[0,T]}}.
\end{gathered}
\end{equation}
It defines an upper bound on the worst-case amplification of the system over all $\Delta \in IQC(\Psi, M)$ for inputs $d(t) \in L_2[0,T]$ and the respective finite time horizon $[0, T]$ with $T \in (0, \infty)$.

\subsection{Bounded Real Lemma for LTV Systems including IQCs}\label{ss:IQCLemma}
A dissipation inequality using the extended system $G$ (\ref{eq:Pext}) and the finite time horizon IQC (\ref{eq:iqctd1}) is formulated to bound either the worst-case gain in (\ref{eq:E2PWC}) or (\ref{eq:L2WC}) of the interconnection $F_u(G_t, \Delta)$ (\cite{Seiler2019, BiePfi2018}). The respective dissipation inequality can be expressed as an equivalent RDE leading to the following two Theorems:
\begin{mytheo}\label{thm:E2EuclGain}
Let $F_u(G_t, \Delta)$ be well posed $\forall \, \Delta \in IQC(\Psi, M)$, 
then $\|F_u(G_t,\Delta)\|_2 < \gamma$ if there exist a
continuously differentiable $P: \R^+_0 \rightarrow \mathbb{S}^{n_x}$ such that
\begin{equation}
\label{eq:Pcondition_E2P}
P(T) = \frac{1}{\gamma}C_2(T)^TC_2(T),
\end{equation}
\begin{equation}
\label{eq:classicRDE_E2P}
\begin{split}
\dot P = Q + P\tilde A+\tilde A^TP-PSP \qquad \forall t \in [0,T]
\end{split}
\end{equation}
and
\begin{equation}
  \label{Rinv_E2P}
  \begin{split}
R = \bsmtx D_{11}^TMD_{11} & D_{11}^TMD_{12}\\
 D_{12}^TMD_{11}   & D_{12}^TMD_{12} - \gamma I_{n_d}\esmtx <0,
\end{split}
\end{equation}
with
\begin{equation}
\label{eq:Atilde_E2P}
\begin{split}
\tilde A = \bsmtx B_1 & B_2\esmtx R^{-1} \bsmtx (C_1^TMD_{11} )^T \\ (C_1^TMD_{12})^T \esmtx - A,
\end{split}
\end{equation}
\begin{equation}
\label{eq:S_E2P}
\begin{split}
S = -\bsmtx B_1 & B_2 \esmtx R^{-1} \bsmtx B_1^T \\ B_2^T \esmtx,
\end{split}
\end{equation}
\begin{equation}
\label{eq:Q_E2P}
\begin{split}
Q =& - C_1^TMC_1
+\bsmtx (C_1^TMD_{11} \\ (C_1^TMD_{12} \esmtx^T R^{-1} \bsmtx (C_1^TMD_{11})^T \\ (C_1^TMD_{12} )^T\esmtx.
\end{split}
\end{equation}
\end{mytheo}
\
\begin{IEEEproof}
The proof is only sketched and a detailed version is given in \cite{Seiler2019}. It is based on the definition of a time-dependent quadratic storage function
$V: \R^{n_x}\times \R^+_0 \rightarrow \R^+_0$. After perturbing (\ref{eq:classicRDE_E2P}), the resulting Riccati inequality
can be rewritten as an LMI applying the Schur complement. The equivalence is guaranteed by condition (\ref{Rinv_E2P}), which also ensures the invertability of $R$.
Multiplying $[x^T ,w^T ,d^T ]$ and $[ x^T ,w^T, d^T ]^T$ 
on the left and right side respectively of the LMI results in a dissipation inequality.
Integration from $0$ to $T$ for zero initial conditions gives
\begin{equation}
\label{eq:E2Pproof1}
x(T)^TP(T)x(T) - \gamma \int_0^T{d(t)^Td(t)dt} + \int_0^T{z(t)^TMz(t)dt} \le 0,
\end{equation}
where the last term can be neglected according to (\ref{eq:iqctd1}).
Equality (\ref{eq:Pcondition_E2P}) is perturbed and left and right multiplied with $x(T)^T$ and $x(T)$ respectively
resulting in
\begin{equation}
\label{eq:E2Pproof2}
\begin{split}
x(T)^TP(T)x(T)-\frac{1}{\gamma}x(T)^TC(T)^TC(T)x(T) \\= x(T)^TP(T)x(T)-\frac{1}{\gamma}e(T)^Te(T)\ge 0.
\end{split}
\end{equation}
After applying $\Delta \in IQC(\Psi, M)$, (\ref{eq:E2Pproof1}) is substituted in (\ref{eq:E2Pproof2}). Subsequently,   the vector 2-norm (Euclidean vector norm) $\norm{e(T)}_2^2 = e(T)^Te(T)$ is used to conclude that, the upper bound on (\ref{eq:E2PWC}) is given by $\gamma$.
\end{IEEEproof}

\begin{mytheo}\label{thm:L2gain}
Let $F_u(G_t, \Delta)$ be well posed $\forall \, \Delta \in IQC(\Psi, M)$, 
then $\|F_u(G_t,\Delta)\|_{2[0,T]} < \gamma$ if there exist a
continuously differentiable $P: \R^+_0 \rightarrow \mathbb{S}^{n_x}$ such that
\begin{equation}
\label{eq:Pcondition_L2}
P(T) = 0,
\end{equation}
\begin{equation}
\label{eq:classicRDE}
\begin{split}
\dot P = \hat{Q} + P\hat{A}+ \hat{A}^TP-P\hat{S}P \qquad \forall t \in [0,T]
\end{split}
\end{equation}
and
\begin{equation}
  \label{Rinv}
  \begin{split}
\hat{R} = \bsmtx D_{11}^TMD_{11} + D_{21}^TD_{21} & D_{11}^TMD_{12} + D_{21}^TD_{22}\\
 D_{12}^TMD_{11} + D_{22}^TD_{21}   & D_{12}^TMD_{12} + D_{22}^TD_{22} - \gamma^2I_{n_d}\esmtx <0,
\end{split}
\end{equation}
with
\begin{equation}
\label{eq:Atilde}
\begin{split}
\hat{A} = \bsmtx B_1 & B_2\esmtx \hat{R}^{-1} \bsmtx (C_1^TMD_{11} +C_2^TD_{21})^T \\ (C_1^TMD_{12} + C_2^TD_{22})^T \esmtx - A,
\end{split}
\end{equation}
\begin{equation}
\label{eq:S}
\begin{split}
\hat{S} = -\bsmtx B_1 & B_2 \esmtx \hat{R}^{-1} \bsmtx B_1^T \\ B_2^T \esmtx,
\end{split}
\end{equation}
\begin{equation}
\label{eq:Q}
\begin{split}
\hat{Q} =& - C_1^TMC_1 - C_2^TC_2 \\
&+\bsmtx (C_1^TMD_{11} +C_2^TD_{21})^T \\ (C_1^TMD_{12} + C_2^TD_{22})^T \esmtx^T \hat{R}^{-1} 
\bsmtx (C_1^TMD_{11} +C_2^TD_{21})^T \\ (C_1^TMD_{12} + C_2^TD_{22})^T\esmtx.
\end{split}
\end{equation}
\end{mytheo}
\begin{IEEEproof}
The proof follows a similar structure as for Theorem 1 as it is again based on the definition of a time-dependent quadratic storage function
$V: \R^{n_x}\times \R^+_0 \rightarrow \R^+_0$. After perturbing (\ref{eq:classicRDE}) the resulting Riccati inequality
can be rewritten as an LMI applying the Schur complement.
Multiplying $\bsmtx x^T ,& w^T ,&d^T \esmtx$ and $\bsmtx x^T ,& w^T, &d^T \esmtx^T$ 
on the left/right side respectively of the LMI results in a dissipation inequality. The integration provides
the upper bound $\gamma$ implied by $\norm{e(t)}_{2[0,T]}\le \gamma \norm{d(t)}_{2[0,T]}$ after applying the final condition $P(T)$ and the IQC conditions.
\end{IEEEproof}

\section{Computational Approach}
\label{sec:CompAppr}
The results of Section \ref{ss:IQCLemma} have yet to be transformed into a
feasible computational optimization problem. In general, a given uncertainty $\Delta$ can
be represented by an infinite set of IQCs. The selection of a fixed set of filters $\Psi$ and a free parametrization of $M$ can be found frequently in literature e.g., \cite{PfiSei2} and \cite{Veenman2016}. This means, $M$ lies within a feasibility set $\mathcal{M}$ such that $\Delta \in IQC(\Psi, M)$ for all $M\in \mathcal{M}$.
Feasible parameterizations for LTI real diagonally repeated as well as LTI dynamic uncertainties can be found in \cite{Veenman2016}. The former is described by Example \ref{ex:delta} and the latter by Example \ref{ex:Delta}.
\begin{exmp}\label{ex:delta}
Let $\Delta = \delta I_{n_v}$ be a LTI real diagonally $n_v$ repeated parametric uncertainty $\delta$, with $\delta \in \R$ and $|\delta| \le b$, with $b \in \R$. A valid time domain IQC for $\Delta$
is defined by $\Psi = \bsmtx \psi_\nu \otimes I_{n_v} & 0 \\ 0 & \psi_\nu \otimes I_{n_v} \esmtx$ and $\mathcal{M} := \{ M = \bsmtx b^2X & Y \\ Y^T & -X\esmtx : X = X^T > 0 \in \mathbb{S}^{n_v(\nu+1)}, Y = -Y^T \in \R^{n_v(\nu+1) \times n_v(\nu+1)}\}$.
A typical choice for $\psi_\nu \in \RH^{(\nu+1) \times 1}$ is: 
\begin{equation}\label{eq:psi_nu}
\psi_\nu = \bmtx 1 & \frac{s+\rho}{s-\rho}& \dots & \frac{(s+\rho)^\nu}{{(s-\rho)}^\nu} \emtx^T\, , \, \rho < 0\, , \, \nu \in \mathbb{N}_0.
\end{equation}
\end{exmp}
\begin{exmp}\label{ex:Delta}
Let $\Delta$ be a LTI dynamic uncertainty, with $\Delta \in \RH$ and $\norm{\Delta}_{\infty} \le b$. A valid time domain IQC for $\Delta$
is defined by $\Psi = \bsmtx \psi_\nu \otimes I_{n_v} & 0 \\ 0 & \psi_\nu \otimes I_{n_v} \esmtx$ and $\mathcal{M} := \{ M = \bsmtx b^2X & 0 \\ 0 & -X\esmtx : X = X^T > 0 \in \mathbb{S}^{n_v(\nu+1)}\}$.
A typical choice for $\psi_\nu \in \RH^{(\nu+1) \times 1}$ is: 
\begin{equation}\label{eq:psi_nu}
\psi_\nu = \bmtx 1 & \frac{s+\rho}{s-\rho}& \dots & \frac{(s+\rho)^\nu}{{(s-\rho)}^\nu} \emtx^T\, , \, \rho < 0\, , \, \nu \in \mathbb{N}_0.
\end{equation}
\end{exmp}
In both examples the matrix variables $X$ and $Y$ are free parameters, whereas $ \psi_\nu $ is a fixed basis function with preselected $\nu$ and $\rho$.

Consequently, the upper bound on $\gamma$ on the worst case gain in Theorem \ref{thm:E2EuclGain} and Theorem \ref{thm:L2gain} depends on the choice of the IQC parameterization. Hence, an optimization problem over the feasible set of IQC parameterizations $M \in\mathcal{M}$ minimizing $\gamma$ constrained by the integrability of the RDE can be derived. In case of the finite horizon worst case $L_2[0,T]$ to $\norm{e(T)}_2$ gain, it is written as:
\begin{flalign*}
&\qquad \min_{\substack{M \in \mathcal{M}}} \gamma&\\
&\qquad \mbox{such that $\forall t \in [0,T]$}&
\end{flalign*}
\begin{equation}
\label{OptiProblem}
\begin{split}
&P(T) = \frac{1}{\gamma}C_2(T)^TC_2(T) \\
&\dot P = Q + P\tilde A+\tilde A^TP-PSP \\
& R<0.
\end{split}
\end{equation}
Note that $M$ enters (\ref{OptiProblem}) in a non-convex way.
The nonlinear optimization problem for the finite horizon worst case $L_2[0,T]$ gain can easily be derived from (\ref{OptiProblem}) by replacing $\tilde{A}$, $S$ and $Q$ with $\hat{A}$, $\hat{S}$ and $\hat{Q}$ respectively as well as changing the final condition to $P(T)=0$.

\subsection{Algorithm}

A novel, custom-tailored optimization algorithm is proposed to deal with problem (\ref{OptiProblem}) efficiently.
The optimization essentially consists of a simple bisection nested within a global optimization algorithm. 
The bisection is used to obtain a minimal $\gamma$ for a given $M$, i.e. bisect (\ref{OptiProblem}) with a fixed $M$.
The optimization over $M \in \mathcal{M}$ is based on the L-SHADE proposed in \cite{Tanabe2014}, but it is significantly tailored towards the specific problem (\ref{OptiProblem}). The L-SHADE algorithm belongs to the class of so-called meta-heuristics. Its key philosophy can be summarized in the following way: 

At the start of the optimization, a random set of possible tuning parameters (populations) is generated, i.e. specifically for (\ref{OptiProblem}) an initial set $\{M_j^{(1)}\}_{j=1}^{n_\text{P}}$, where $n_\text{P}$ is the number of populations. Throughout this section, $M_j^{(i)}$ is the elements of the IQC parameterization stacked as a vector, where $j$ denotes the $j^{\textrm{th}}$ individual within the population and $i$ the population iteration. At each new iteration $i$, the population is updated with a bias towards the best 10\% solutions of the previous iteration $i-1$ by:
\begin{equation}
	\label{eq:MutationOp}
	{M_j}^{(i)} = M_j^{(i-1)} +F_{j} (M_{\text{pbest},j}^{(i-1)} - M^{(i-1)}_j +M_{\text{r1},j}^{(i-1)} - M_{\text{r2},j}^{((i-1)}).
\end{equation}
In (\ref{eq:MutationOp}), $M_\text{pbest}$ is a randomly selected individual from the best 10\% of the population, $M_\text{r1}$ and  $M_\text{r2}$ are two randomly selected individuals from the whole population, and $F_j$ is a scaling factor chosen as a Cauchy distributed random scalar with variance $0.1$ and a randomly selected mean value out of the set $S_F$ of previously well-performing scaling factors (see \cite{Tanabe2014} for details). 

After performing the mutation, each element in $M_j^{(i)}$ is potentially replaced with the respective element of its parent $M_j^{(i-1)}$ using binomial crossover.A uniformly distributed random number in the interval $(0,1)$ is assigned to each element in $M_j^{(i)}$. If this number is larger than the element's crossover rate ($CR$), the respective element is replaced by its parent. The crossover rate is chosen as  a normal distributed random number with variance $0.1$ and a mean value randomly chosen from the set $S_{CR}$ of previously well-performing crossover rates.
However, one random element in $M_j^{(i)}$ always remains updated independently of its crossover rate.

After finishing the crossover, the bound constraints are checked. In case of a violation, the respective elements in $M_j^{(i)}$  are set to the arithmetic mean value of the corresponding parental elements in $M_{j}^{(i-1)}$ and the respective violated boundary.

The following adaptations have been made to the original algorithm proposed in \cite{Tanabe2014}:
The algorithm uses a logarithmic scaling of the decision variables instead of the linear scaling used in \cite{Tanabe2014}.  The search space for the IQC parameters usually covers several order of magnitudes with no clear indication for good initial values. For instance, the diagonal entries of $X$ in Example \ref{ex:Delta} are only restricted by positiveness. By searching over a logarithmic scale, the correlation between a change in $\gamma$ and the variation of the elements in the IQC parameterization $M$ is better represented.
This is especially true considering very small magnitudes. Hence, the meta-heuristic converges at the same speed independently of the optimal solution's magnitudes.
Using a logarithmic search space, the single elements in the decision vector $M_{j}^{(i)}$ are now represented by two elements containing its sign and exponent to base ten. These two elements are stacked into single vectors $M_{\text{Log},j}^{(i)}$ forming the actual set of tuning parameters in the logarithmic space $\{M_{\text{Log},j}^{(1)}\}_{j=1}^{n_\text{P}}$.
Note that this logarithmic scaling effectively increases the number of decision variables. It should be emphasized that, in general, it does not double the number of variables as many IQC parameters are sign defined, e.g. the diagonal entries of $X$ in Example \ref{ex:Delta}.

MHs do not require a valid initial population set, i.e. that a finite $\gamma$ value exists for a given $M_j^{(i)}$ such that the RDE in (\ref{OptiProblem}) is fully integrable. In general, $\gamma$ cannot be calculated for all $M\in \mathcal{M}$ due to the RDE's finite escape time \cite{AbouKandil2003}. The algorithm is adjusted to require at least $20\%$ valid members in its initial population before commencing the mutation to improve convergence. In the case of multiple IQCs, an optional downscaling of the uncertainties' norm bounds by a factor $k_\text{IQC}$ facilitates identifying a valid initial population by increasing the feasible search space. Note that optima locations are not noticeably affected by the norm bound. Thus, procedural rescaling does not adversely affect the search performance. Furthermore, initial guesses $M_{\text{Log,init}}$ can be added to the initial population.

Integrating the RDE is the main contributor to the algorithm's computation time. Hence, several adjustments have been made to reduce the number of necessary integrations. In general, narrow bisection bounds are beneficial. The lower bound is provided either by the nominal $\gamma$ value \cite{Green1995} or a theoretical lower bound imposed by $R<0$ and Schur's complement.
For the initial population, an adaptive upper bound is implemented. As long as the RDE is not entirely solvable, $\gamma_\text{UB}$ increases iteratively until a maximum value of $10^{20}$ is reached.
Inside the iterative search procedure, the upper bound used for $M_j^{(i)}$ is the minimal $\gamma_j^{(i-1)}$ identified for the corresponding parent $M_j^{(i-1)}$. If the RDE is not fully solvable for this upper bound, no bisection is executed, and $M_j^{(i)}$ gets assigned $\gamma_j^{(i)}  = 10^{20}$. This adjustment is based on the fact that subsequent mutations are only influenced by offspring whose $\gamma$ value is lower than the respective parent's.

Finally, the condition $R < 0$ (\ref{OptiProblem}) can be further exploited to reduce the number of integrations in the optimization. 
If in a bisection step $R\ge 0$, the present $\gamma$ value is treated as a failed integration. The integration is also skipped if $R$'s is poorly conditioned, i.e. its condition number is larger than a user-defined threshold. Hence, numerical difficulties integrating the RDE are avoided. Extensive test scenarios showed no adverse effects on the search performance.

In general, RDEs are considered stiff \cite{AbouKandil2003}. Hence, the RDE is solved via Matlab's built-in solver \texttt{ODE15s}. Its integrated event function is used to detect blow-ups resulting from finite-escape times shorter than the analysis horizon and terminate the integration. The event function triggers if the largest absolute eigenvalue of $\dot{P}$ is above a provided threshold.
The computational effort is further reduced by exploiting the RDE's symmetry. Thus, only $0.5n(n+1)$ instead of $n^2$ equations must be solved.

\begin{algorithm}[ht!]
\caption{Log-L-SHADE }
\label{Alg:Log-L-SHADE}
\begin{algorithmic}[1]

\State {\textbf{Input:} $n_\text{P,max}$, $n_\text{P,min}$, $i_\text{max}$, $k_{CR}$,  $k_{F}$, $G$, $\mathcal{M}$, $s_\text{min}$, $s_\text{max}$, $\gamma_\text{lim}$, $M_\text{Log}^{(0)}$, $k_\text{IQC}$, $\gamma_\text{LB}$, $\gamma_\text{UB}$, $\epsilon_\text{BS}$}
\State{\textbf{Output:} $\gamma_\text{best}$, $M_\text{Log,best}$}
\State{\textbf{Initialize:} $S_F$, $S_{CR}$}
\State{Scale original IQC norm bound by $k_\text{IQC}$}
\While{Less than $20\%$ valid initial members}
\State{Generate random initial population $\{M_{\text{log},j}^{(1)}\}_{j=1}^{n_\text{P}}$}
\State{Calculate $\gamma(M_{\text{Log},j}^{(i)})$ via bisection}
\EndWhile
\State{Find current best solution $M_\text{Log,best}$ and fitness $\gamma_\text{best}$}
\State{Set IQC norm bound upscaling threshold $n_\text{P,IQC}=n_\text{P,max}$}
	
\While{($i \le i_\text{max}$ OR $\gamma_\text{best}>\gamma_\text{lim}$) AND $k_\text{IQC}<1$}
	\State{$i = i+1$}
	\If{($\gamma_\text{best} \le \gamma_\text{lim}$ OR $n_{P}<0.8 n_\text{P,IQC}$) AND \newline
	\hspace*{1.54em}$k_\text{IQC}<1$}
	\State{Set $n_\text{P,IQC} = n_\text{P}$ and $k_\text{IQC}  = \text{min}(3k_\text{IQC}, 1)$}
	\State{Upscale norm bound, recalculate $\gamma(M_{\text{Log},j}^{(i-1)})$ with\newline
	\hspace*{3.1em}user $\gamma_\text{UB}$, and update $M_\text{Log,best}$ and $\gamma_\text{best}$}
	\EndIf
	\For{$j =1$ to $n_\text{P}$}
		\State{Calculate $M_{\text{Log},j}^{(i)}$ using (\ref{eq:MutationOp}) and crossover, \newline
		\hspace*{3.0em}enforce boundaries, and  ${M}^{(i)}_{\text{Log},j} \in \mathcal{M}$}
		\If{RDE is solvable for ${M}^{(i)}_{\text{Log},j}$ and $\gamma({M}^{(i-1)}_{\text{Log},j})$}
		\State{Execute bisection with $\gamma_\text{UB} = \gamma(M^{(i-1)}_{\text{Log},j})$}
		\Else
			\State{Skip bisection and set $\gamma(M_\text{Log,j}^{(i)}) = 10^{20}$}
		\EndIf
	\If{$\gamma(M_{\text{Log},j}^{(i)})>\gamma(M_{\text{Log},j}^{(i-1)})$}
		{$M_{\text{Log},j}^{(i)}=M_{\text{Log},j}^{(i-1)}$} 
	\EndIf
	\EndFor
	\State{Update $S_F$ and $S_{CR}$ based on successful $F$ and $CR$}
	\State{Identify current best solution $M_\text{Log,best}$ and fitness $\gamma_\text{best}$}
	\State{Update population $n_\text{P}$ size via (\ref{eq:PopSizeUp}) and remove worst \newline
	\hspace*{1.54em}solutions from $\{M_{\text{Log},j}^{(i)}\}_{j=1}^{n_\text{P}}$}
\EndWhile

\end{algorithmic}

\end{algorithm}

Algorithm 1 presents pseudo-code to illustrate the implementation of the optimization problem.
The user must provide a total of sixteen inputs. The first two are the maximum and the minimum number of  populations $n_\text{P,max}$ and $n_\text{P,min}$, respectively. These are followed by the maximum amount of population iterations $i_\text{max}$ and the number of successful crossover rates and scaling factors $k_{CR}$ and $k_{F}$. 
$G$ is the extended system in (\ref{eq:Pext}), which includes the user-selected fixed IQC filter $\Psi$. It is followed by $\mathcal{M}$ describing the set of feasible IQC parameterizations. 
The inputs $s_\text{min}$ and $s_\text{max}$, with $s_\text{min}\in \field{N}^{n_m}$ and $s_\text{max} \in \field{N}^{n_m}$, define the logarithmic search space, i.e. minimal and maximal exponent, for each element in $M_j^{(i)}$.
Furthermore, the optional norm bound scaling $k_\text{IQC}$ and initial guess  $M_\text{Log,init}$. The input $\gamma_\text{lim}$ is used as scaling as well as a terminal condition. The remaining three inputs are required to run the bisection, its lower and upper bound  $\gamma_\text{LB}$ and $\gamma_\text{UB}$, as well as its relative tolerance $\epsilon_\text{BS}$.

The algorithm is initialized with the vectors $S_F\in \R^{k_F}$ and $S_{CR}\in \R^{k_{CR}}$ containing $k_{F}$ and $k_{CR}$ elements, respectively, with a value of $0.5$. 
Subsequently, an initial population is generated and evaluated via bisection using the user-defined bounds. After the required amount of valid members is reached,
the current best solution $M_\text{Log,best}$ and respective $\gamma_\text{best}$ are identified.

Now, the tailored MH's iterative search procedure starts. 
First, the uncertainty norm bounds are upscaled if either $\gamma_\text{best}<\gamma_\text{lim}$ or  $n_{P}<0.8n_\text{P,IQC}$, with $n_\text{P,IQC}=n_\text{P,max}$ for the first iteration and $k_\text{IQC}<1$. If the norm bound is upscaled, the population is reevaluated and the new best solution identified.
It proceeds with the population updated using mutation (\ref{eq:MutationOp}) and crossover in the logarithmic domain.
Afterwards, the bisection with updated upper bounds is executed to calculate the minimal $\gamma({M}_{\text{Log},j}^{(i)})$ to a relative accuracy of $\epsilon_\text{BS}$. The computation is fully parallelizable, i.e. the number of accessible workers/processor cores is directly inverse to the computation time. 
If the child is an improvement over its parent, it replaces its parent in the next population iteration. Otherwise, the parent is used for the next iteration.

Before the next iteration starts, the population size $n_\text{P}$ is updated using:
\begin{equation}\label{eq:PopSizeUp}
n_\text{P} = n_\text{P,max} - \text{round}\left(\frac{(n_\text{P,max} - n_\text{P,min})}{i_\text{max}n_\text{P,max}}i \right).
\end{equation}
If it decreases, then the worst excess solutions in $\{M_{\text{Log},j}^{(1)}\}_{j=1}^{n_\text{P}}$ are removed to match the new population size. 
The optimization concludes as soon as the maximum number of iterations is reached or the best $\gamma$ value is lower than the threshold given that the norm bounds are rescaled. The algorithm returns $\gamma_\text{best}$ and the respective solution $M_\text{Log,best}$.


\section{Example}\label{sec:Example}
The algorithm proposed in Section \ref{sec:CompAppr} is applied to identify the worst-case aerodynamic load acting on a space launcher during its atmospheric ascent.  It covers the flight segment from $25\text{s}$ to $95\text{s}$ after lift-off, which includes the most critical regions of atmospheric disturbance and the dynamic pressure acting on the launcher.
A realistic uncertainty set modeled via IQCs as well as a realistic wind disturbance model are introduced. The LTV worst-case results are compared a Monte Carlo type analysis conducted on the nonlinear model covering the same uncertainty and disturbance set. 

\subsection{Space Launcher Model}\label{sec:LauncherModel}
The investigated ELV is built of three solid rocket stages and an upper module using liquid propellant. It is designed to launch small payloads into polar and low earth orbits.
During the ascent, the ELV is exposed to high dynamic pressures leading to substantial aerodynamic loads. This is accompanied by unsteady aerodynamics in the transonic region \cite{Blevins2014}. The launcher is also subject to various disturbances. The most influential of these is wind \cite{Johnson2008}. Nevertheless, the launcher has to stay inside a small design envelope to maintain its structural integrity and to deliver the payload into the correct injection orbit. 

\subsubsection{Nonlinear Dynamics}
\label{ss:NonLinDyn}

The ELV is assumed symmetrical during the ascent as it is common practice for most space launchers \cite{Orr2009}. Thus, the pitch and yaw motion can be described by the same dynamics with negligible
cross-coupling \cite{Greensite67I}.
As a result, it is sufficient to solely consider the space launcher's pitch motion for the analysis.
Further, following standard practice, the influences of a spherical and rotating earth to analyze the atmospheric flight phase are neglected \cite{Greensite67I}.
Additionally, propellant sloshing and nozzle inertia are ignored. These mainly influence the flexible modes of an ELV. However, the considered analysis aims at finding the worst-case static loads.

\begin{figure}[htb!]
\centering
\def\svgwidth{\linewidth}
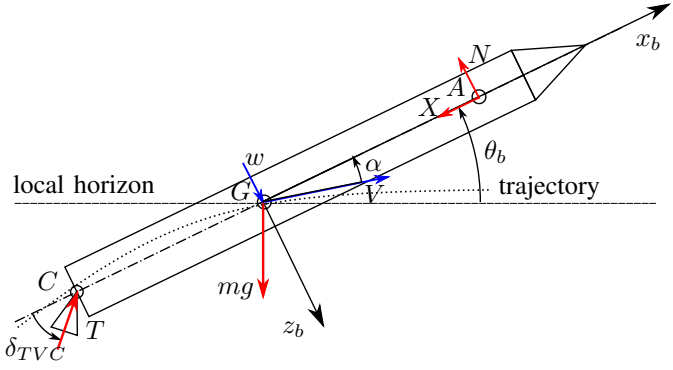
\caption{Launcher vehicle and trajectory frame dynamics}
\label{fig:Launcher}
\end{figure}

In Fig. \ref{fig:Launcher}, a schematic of the launch vehicle is given. A launcher-fixed coordinate system with subscript $b$ is used to formulate the nonlinear equations of motion. It is fixed to the center of mass of the launcher $G$. Its $x_b$-axis is aligned with the launcher symmetry axis defined in the direction of forward travel. The $z_b$-axis forms a right-hand system with the $y_b$-axis pointing out of the page.
Accordingly, the rigid body motion in the pitch plane formulated in body fixed coordinates is described by
\begin{equation}
\label{eq:PitchNonLin}
\begin{split}
\ddot{\theta}_b(t) &= \frac{\sum{M_y(Ma,\alpha, h,t)}}{J_y(t)} \\
&= \frac{N(Ma,\alpha,h,t) l_{GA}(t,Ma)}{J_y(t)} - \frac{T(t) l_{CG}(t)}{J_y(t)}\sin{\delta_\text{TVC}(t)}\\
\ddot{x}_b(t) &=  \frac{\sum{F_x(Ma,\alpha,h,t)}}{m(t)} - \dot{\theta}_b(t)\dot{z}_b(t) \\&= \frac{T(t)\cos{\delta_\text{TVC}(t)}-X(Ma,\alpha,h ,t)}{m(t)} \\
&- g_0(h) \sin{\theta_b}(t) - \dot{\theta}_b(t)\dot{z}_b(t)\\
\ddot{z}_b(t) &=  \frac{\sum{F_z(Ma,\alpha, h, t)}}{m(t)} + \dot{\theta}_b(t)\dot{x}_b(t) \\
&= -\frac{N(Ma,\alpha,h,t)}{m(t)} - \frac{T(t)}{m(t)}\sin{\delta_\text{TVC}(t)}\\
&+ g_0(h) \cos{\theta_b(t)}+ \dot{\theta}_b(t)\dot{x}_b(t),
\end{split}
\end{equation}
where $\sum M_y$ is the sum of the angular moments in the pitch plane with respect to the center of gravity $G$.
$\sum F_x$ and $\sum F_z$  describe the sum of forces in $x_b$ and $z_b$-direction, respectively.
The angle $\theta_b$ is the pitch angle of the launcher describing the angle between the body axis and the local horizon. 
The normal aerodynamic force is denoted $N$. It is described by
\begin{equation}
\label{eq:lift}
N(Ma,\alpha,h,t) = Q(h,t) S_{ref} C_{N_{\alpha}}(Ma) \alpha(t),
\end{equation}
with 
\begin{equation}
\label{eq:DynPres}
Q(h, t) = 0.5 \rho(h, t) V(t)^2
\end{equation}
being the dynamic pressure. $C_{N_\alpha}$ is the normal lift force coefficient, which depends on the Mach number $Ma$. 
The density of the air $\rho$ is calculated according to the international standard atmosphere (ISA).
$N$ acts parallel to the $z_b$-axis. It is defined positive in negative $z_b$ direction. The axial aerodynamic force $X$ is defined in the same way but with respect to the $x_b$-axis.
$X$ is described by the equation
\begin{equation}\label{eq:drag}
X(Ma,\alpha, h, t) = Q(h,t) S_\text{ref} (C_{X_0}(Ma) + C_{X_\alpha}(Ma)\alpha),
\end{equation}
where $C_{x_0}$ is the zero-lift and $C_{x_\alpha}$ lift dependent axial force coefficient, respectively. Both coefficients are $Ma$ dependent. 
The definition differs from the common aerospace convention formulating lift and drag parallel respectively orthogonal to the aerodynamic velocity $V$.
In (\ref{eq:lift}) and (\ref{eq:drag}), the angle of attack is approximated as
\begin{equation}
\label{eq:alpha}
\alpha(t) \approx \frac{\dot{z}_b(t) - v_w(t)}{\dot{x}_b(t)},
\end{equation}
where $v_w$ is the wind speed. It is aligned with the $z_b$-axis and defined as positive in $z_b$-direction.
$T$ denotes the time-dependent thrust of the engine acting at the nozzle pivot point $C$.
The geometric variables $l_{GA}$ and $l_{CG}$ denote the distance between the center of gravity $G$ and the center of aerodynamic pressure $A$ and $C$, respectively. All aerodynamic forces act at on $A$.
$G$ moves forward during the flight due to the propellant burn, whereas $A$'s location depends on the $Ma$ number.
$J_y$ and $m$ denote the mass moment of inertia and the launcher's mass, respectively, which vary with time due to the fuel burn.
The gravitational acceleration $g_0$ is calculated according to the world geodetic systems WGS-84 (\cite{WGS84}) as a function of altitude.
As the only control variable, the deflection of the thrust vectoring control (TVC) $\delta_\text{TVC}$ is available.
The dynamics of the thrust vectoring control are given by the following second order system:
\begin{equation}\label{eq:TVC}
G_\text{TVC} = \frac{1}{0.000374 s^2 + 0.0384s + 1}.
\end{equation}

\


\subsubsection{Trajectory and Control Design}
\label{ss:PerfMeas}

The ascent trajectory is a standard gravity turn assuring $\alpha\approx 0$  and $\delta_{TVC}\approx 0$ during the nominal ascent.
The purpose is to minimize the static aerodynamic load and maximize the longitudinal acceleration. This is beneficial for the launcher design as it minimizes structural and fuel mass. The reference flight path parameters for the analyzed ELV are calculated by iteratively solving the following initial value problem
\begin{equation}
  \label{eq:GravityTurn}
  \begin{split}
   \dot{h}_\text{ref}(t) &=\dot{x}_{b,\text{ref}}(t) \sin{\theta_{b,\text{ref}}(t)}\\
    \dot{\theta}_{b,\text{ref}}(t) &= -\frac{g_0(h)}{\dot{x}_{b,ref}(t)}\cos{\theta_{b,\text{ref}}(t)}\\
   \ddot{x}_{b,\text{ref}}(t) &= \frac{T(t)-X(Ma, \alpha, h,t)}{m} - g_0(h)\sin{\theta_{b,\text{ref}}(h)}
  \end{split}
\end{equation}
given in \cite{Wiesel2010}.
The initial values for $\theta_{b}$, $\dot{x}_{b}$, and the altitude $h$ are iterated until the trajectory's desired terminal values are achieved.
The calculated flight path represents a common mission scenario \cite{Gallucci2012}.

Due to the lack of aerodynamic surfaces, the analyzed ELV is aerodynamically unstable. Hence, feedback control is required to stabilize its dynamics along the trajectory. Furthermore, it is necessary to track the pre-calculated gravity turn trajectory. In this paper, this is achieved by executing a pitch program following the reference pitch angle 
$\theta_{b,ref}$ calculated via (\ref{eq:GravityTurn}).

Many modern launch vehicles still rely on rather basic proportional, integral, and derivative (PID) control \cite{Orr2009}. 
Therefore, a simple fixed-gain PID controller $C$ with a derivative filter for $\theta_t$ tracking was designed.
As the design point, the point of the maximum dynamic pressure along the reference trajectory was chosen, which is at $t =54\text{s}$.
The gains are calculated via loop shaping, providing a maximum tracking bandwidth of $6rad/s$ without actuators.  The controller satisfies phase and gain margins of $45^\circ$ and $6\text{dB}$ respectively along the trajectory as proposed in \cite{Greensite67VII}.

The controller was evaluated in the nominal nonlinear ELV simulation. The ELV follows the pitch program precisely under nominal conditions ($|\Delta\theta(t)| < 0.00013^\circ$). Therefore, $|\alpha | \approx 0^\circ$ without wind disturbances.
Under allowable wind disturbance, i.e. suitable launch conditions, $|Q(t)\alpha(t)| < 22000\text{Pa}^\circ$ as it is required by ESA guidelines. This assures that the ELVs structural limit loads are not exceeded \cite{Vega1}.


\

\subsubsection{Linear Dynamics}
\label{ss:LinDyn}
The nonlinear launcher dynamics in (\ref{eq:PitchNonLin}) are linearized along the ascent trajectory. 
This reduces the various parameter dependencies in (\ref{eq:PitchNonLin}) to a sole time dependence along the calculated trajectory.
Thus, the linearization results in a (general) LTV system $G_t$ as described by (\ref{eq:LTVsys}) in Section \ref{sec:RobAnalysis}.
In case of the analyzed ELV, the input vector is $u(t) = [\delta_{TVC}, w]^T$, the output vector is $y(t) = [\theta_b, Q\alpha]^T$, and the states are $\theta_b$, $\dot{\theta}_b$ and $\dot{z}_b$. 
Note that $\dot{x}_b$ does not need to be considered in the analysis, as it has no impact on the maximum aerodynamic load.
The matrices $A_{G_t}$, $B_{G_t}$, $C_{G_t}$, and $D_{G_t}$ are calculated via numerical linearization with a step size of $\Delta t = 0.1s$. The chosen grid is dense enough to capture the fast changing dynamics in the transonic region during the ascent. 
\

\subsubsection{Uncertainty Model} \label{ss:UncMod}
The uncertainty in the launcher's aerodynamics arises mainly from the limited means of testing and the resulting reliance on simulation results. Furthermore, the launcher passes through the transonic ($0.8 \le Ma \le 1.2$), for which the estimation of aerodynamic parameters is complicated. Especially difficult to estimate is the center of aerodynamic pressure, due to complicated airflow originating from the payload fairing. This is accounted for by uncertainty in $l_{GA}$, which directly influences the aerodynamic instability of the launcher. In addition, the launcher's center of gravity is subject to uncertainty due to variations in the fuel-burn. Therefore, uncertainty in $l_{CG}$ is introduced. It directly influences the controllability/stabilizability. Finally, higher-order dynamics of the overall system are not explicitly modeled. These are included by uncertainty in the TVC's dynamics. The introduced uncertainties have little influence on the ascent trajectory. Thus, the reference trajectory to obtain the nominal model maintains its validity.

The uncertainties in the launcher's parameters are all described by (repeated) LTI real parametric uncertainties. They are summarized in
Table \ref{tab:UncSet}. 
\begin{table}\label{tab:UncSet}
\caption{Uncertainty set used for the robustness analysis\label{tab:UncSet}}
\centering
\begin{tabular}
{ p{1.3cm} p{1.1cm} p{1.3cm} p{1.6cm}  p{1.1cm} }
 Parameter & Notation & Value & Occurrences & Type\\
 \hline
 $l_{CG}$ 			& $\delta_{l_{CG}}$ 		& $10\%$ 					& $1$ 		& real\\
 $l_{GA}$  			& $\delta_{l_{GA}}$		& $20\%$					& $2$		& real\\
 TVC 				& $\Delta_\text{TVC}$ 			& $|{\frac{2.6s + 14.47}{s + 144.7}}|$	& $1$		&dynamic\\
 \hline
\end{tabular}
\end{table}
The uncertainty in the TVC dynamics is represented with
a dynamic LTI uncertainty $\Delta_\text{TVC}$ with $\norm{\Delta_\text{TVC}}_{\infty}<1$. It is implemented as
\begin{equation}
\label{eq:ActUnc}
G_{TVC} = G_{TVC,nom} (1 + \Delta_\text{TVC}W_\text{TVC}),
\end{equation}
with an weighting filter $W_\text{TVC}(s)$. $W_\text{TVC}$ is calculated based on the approach in \cite{Hindi2002}.
It covers a time delay of $10\text{ms}$ and up to $10\%$ uncertainty in the TVCs static gain, damping ratio, and eigenfrequency.


\

\subsubsection{Wind Model}\label{ss:WindModel}

The wind disturbance is based on the Dryden turbulence filter for light lateral wind turbulence \cite{Hoblit1988}. Dryden turbulence spectra are widely used in aerospace certification processes.

\

\paragraph{Wind Filter Nonlinear Analysis}

The Monte Carlo simulation of the nonlinear model applies the standard Dryden filter for lateral wind turbulence $G_w$:
\begin{equation}
\label{eq:DrydenNonLin}
\begin{split}
\dot{x}_w(t) &= \bmtx 0 & 1 \\ -\left(\frac{V(t)}{L(h)}\right)^2 & -2\frac{V(t)}{L(h)}\emtx x_w(t) 
+ \bmtx 0 \\ \left(\frac{V(h)}{L(h)}\right)^2 \emtx n_w(t)\\
v_w(t) &= \bmtx \sigma(h)\sqrt{\frac{L(h)}{\pi V(t)}} & \sigma(h) \frac{L(h)}{V(t)} \sqrt{\frac{3L(h)}{\pi V(t)}}\emtx x_w(t),
\end{split}
\end{equation}
as given in \cite{Hoblit1988}. In (\ref{eq:DrydenNonLin}), $V(t) = \sqrt{x_b^2 + z_b^2}$ is the velocity of the ELV, $\sigma$ is the turbulence intensity and $L_u$ is the turbulence scale length. 
This filter shapes a white noise input $n_w(t)$ with a power spectral density $\Phi_{n_w} = 1$ into a continuous turbulence signal $v_w(t)$.
Here, the Simulink internal band-limited white noise block is used to generate $n_w(t)$. The lateral filter is chosen, as for altitudes over $533m$ the wind turbulence is defined as being aligned with the body fixed coordinates. This holds for the analyzed trajectory segment. Hence, the calculated wind turbulence is consistent with the definition of $v_w$ in (\ref{eq:alpha}).

\

\paragraph{Wind Filter LTV Analysis}
The wind filter $G_w$ cannot be applied directly in the LTV analysis due to the strict BRL.
Recall that the Dryden filter (\ref{eq:DrydenNonLin}), in contrast, assumes a white noise input. Hence, to get meaningful analysis results, a wind filter for the LTV analysis must be designed to take any bounded L2 signal and generate realistic Dryden turbulence signals. Specifically, the wind filter design goal is to match the power spectral density (PSD) of the Dryden turbulence.

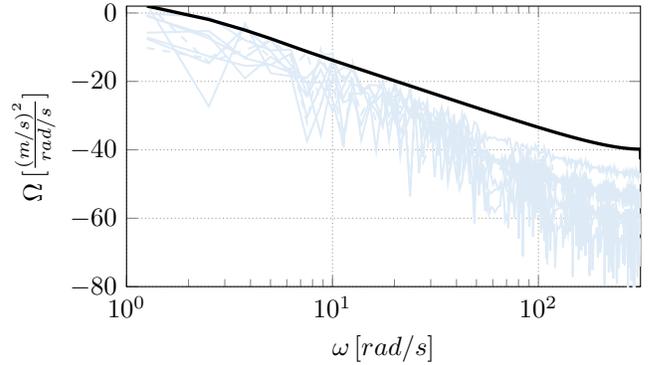
\begin{figure}[ht!]
\centering
\begin{tikzpicture}
\definecolor{blue1}{RGB}{222,235,247}
\definecolor{blue2}{RGB}{158,202,225}
\definecolor{blue3}{RGB}{49,130,189}
\begin{semilogxaxis}[ width = 0.95\columnwidth, height = 0.6\columnwidth, 
  	grid=major, 
   grid style={densely dotted,white!60!black}, 
   xlabel=  $\omega  \,{[rad/s]}$, 		
   ylabel=  $\Omega\,{[\frac{(m/s)^2}{rad/s}]}$, 	
   legend style={at={(0.65,0.97)},anchor=north west},
   legend cell align = {left},
   xmin = 1, xmax = 313, ymin = -80, ymax = 2, 
        ]

\addplot[blue1, line width = 0.75,  no marks] table[x expr = \thisrowno{0} ,y expr = \thisrowno{10} ,col sep=comma] {figures/PSDcompare.csv};\label{Dryden}
\addplot[blue1, line width = 0.75,  no marks] table[x expr = \thisrowno{1} ,y expr = \thisrowno{11} ,col sep=comma] {figures/PSDcompare.csv};

\addplot[blue1, line width = 0.75,  no marks] table[x expr = \thisrowno{2} ,y expr = \thisrowno{12} ,col sep=comma] {figures/PSDcompare.csv};

\addplot[blue1, line width = 0.75,  no marks] table[x expr = \thisrowno{3} ,y expr = \thisrowno{13} ,col sep=comma] {figures/PSDcompare.csv};

\addplot[blue1, line width = 0.75,  no marks] table[x expr = \thisrowno{4} ,y expr = \thisrowno{14} ,col sep=comma] {figures/PSDcompare.csv};

\addplot[blue1, line width = 0.75,  no marks] table[x expr = \thisrowno{5} ,y expr = \thisrowno{15} ,col sep=comma] {figures/PSDcompare.csv};

\addplot[blue1, line width = 0.75,  no marks] table[x expr = \thisrowno{6} ,y expr = \thisrowno{16} ,col sep=comma] {figures/PSDcompare.csv};
\addplot[blue1, dashed,line width = 0.75,  no marks] table[x expr = \thisrowno{7} ,y expr = \thisrowno{17} ,col sep=comma] {figures/PSDcompare.csv};\label{MCBoundUpp}
\addplot[blue1, dashed,line width = 0.75,  no marks] table[x expr = \thisrowno{8} ,y expr = \thisrowno{18} ,col sep=comma] {figures/PSDcompare.csv};

\addplot[black,line width = 1.25,  no marks] table[x expr = \thisrowno{9} ,y expr = \thisrowno{19} ,col sep=comma] {figures/PSDcompare.csv};\label{LTVwind}








\end{semilogxaxis}
\end{tikzpicture}
\caption{Comparison of power spectral densities for the analysis segment from $25\text{s}$ to $30\text{s}$: LTV worst case wind signal (\ref{LTVwind}),  Dryden turbulence (\ref{Dryden})}
\label{fig:PSDcompare}
\end{figure}

The proposed design procedure is as follows: In the first step, $10000$ random wind profiles are generated along the nominal trajectory over the analysis horizon using the Dryden filter (\ref{eq:DrydenNonLin}) at a fixed sampling rate of $100\text{Hz}$. The second step is the calculation of the PSD. The wind signals are divided into $n = 14$ equidistant segments of $5\text{s}$ to account for the varying turbulence intensity along the trajectory.
Subsequently, the PSD $\Omega_{v_{w,i,n}}$ of a segment $n$ of a time-domain wind signal $v_{w,i}(t)$ is calculated using
\begin{equation}\label{eq:FFT}
\Omega_{v_{w,i,n}}(\omega)=\lim_{T\rightarrow \infty}\frac{1}{\pi}\frac{1}{T}\abs{\int_{-T}^{T}v_{w,i}(t)e^{-j\omega t}dt}^2.
\end{equation}
Therefore, the PSD of a time-domain signal is simply the average squared of the signal's Fourier transform. The Fourier transform of the wind signals can be calculated via a fast Fourier transform (FFT). In here,  the internal Matlab function \texttt{fft} was applied. 
This procedure is repeated for all segments $n$ over all wind signals $v_{w,i}(t)$.
In the third step, a transfer function is calculated for each time segment, upper bounding 
the respective $\abs{\Omega_{v_{w,i,n}}(\omega)}$ over all wind signals. The internal Matlab function \texttt{fitmagfrd} is used, calculating a minimum phase first-order transfer function based on log-Chebychev magnitude design. These transfer functions are transformed into consistent state-space models. 
Afterwards, the LTV representation $G_{w,\text{LTV}}$ of the wind model is calculated by linear interpolating the system matrices' coefficients over the analysis horizon.
In the fourth step, the calculated wind filter is evaluated to check if it produces adequate wind signals. Therefore, the nominal launcher model is extended with $G_{w,\text{LTV}}$, and the nominal LTV worst case disturbance signal $d_\text{WC}$ for a given terminal  time is calculated using the approach in \cite{Iannelli2019}.
Filtering $d_\text{WC}$ through $G_{w,\text{LTV}}$ provides the respective worst-case wind signal. Its amplitude is compared to the actual Dryden turbulence.
In case the LTV wind signals underestimate the amplitude of actual Dryden turbulence, steps three and four are repeated with an increased lower bound for \texttt{fitmagfrd} until the amplitudes show an adequate match. In combination, step three and four assure matching PSDs of the LTV worst-case wind signal and actual Dryden turbulence. In Fig. \ref{fig:PSDcompare}, the worst-case LTV wind signal's PSD for the time segment from $25\text{s}$ to $30\text{s}$ is compared to the corresponding Dryden turbulence signal.
The presented procedure to derive the LTV wind filter is not limited to Dryden turbulence but can be easily applied to any available wind data.

\

\subsubsection{Analysis Interconnection}\label{ss:UncMod}
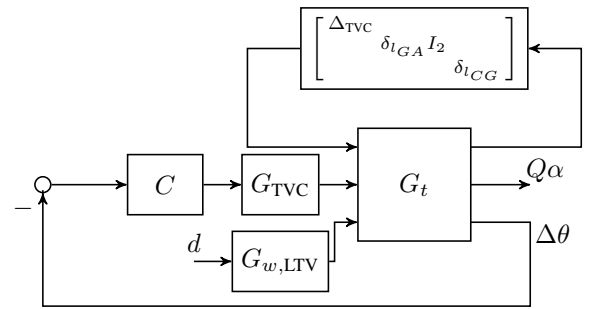
\begin{figure}[ht!]
    \centering
        \begin{tikzpicture}[blockdiag]
	
	
	\node[block, minimum height=1.5cm,minimum width=1.5cm](plant){$G_t$}; 
	\node[block,left = of plant, xshift = +0.0cm, yshift = +0.0cm](TVC) {$ G_\text{TVC}$}; 
	\node[block, left = of TVC](C){$C$};
	\node[sum, left = of C, xshift = -0.5cm](Sum1){};
	\node[block, above = of plant](delta){$\bsmtx \Delta_\text{TVC} &  &  \\  & \delta_{l_{GA}}I_2 &  \\  &  & \delta_{l_{CG}} \esmtx$};
	\node[block, below = of TVC, yshift = +0.3cm](Wind){$G_{w,\text{LTV}}$}; 


\draw[->](TVC.east) -- (plant.west);
\draw[->](C.east) -- (TVC.west);
\draw[->](Sum1.east) -- (C.west);

\draw[->] ([yshift = -0.5cm, xshift =0.08cm]Plant.east) -- node[above,xshift = 0.70cm, yshift = -0.4cm, name=DeltaTheta]{$\Delta{\theta}$}  +(+0.8cm, -0.0cm) |- ($(Sum1.south) +(0cm, -1.5cm)$)  |-
	([yshift = 0]Sum1.south)node[left,yshift =- 0.2cm, name=minus]{$ - $};

\draw[->]([yshift = +0.0cm, xshift = 0.08cm]Plant.east)-- +(0.8cm, +0.0cm)node[above, name = e, xshift = 0.2cm, yshift = -0.1cm]{$Q\alpha$};

\draw[->]([yshift = +0.5cm, xshift = 0.08cm]Plant.east)  -| ($(delta.east) +(0.7cm, 0cm)$) -- (delta.east);
\draw[->](delta.west) --   ($(delta.west) +(-0.7cm, 0cm)$) |- ($(plant.west) +(-0.00cm, 0.5cm)$);

\draw[->](Wind.east) -|  ($(plant.west) +(-0.3cm, -0.50cm)$) -- ($(plant.west) +(-0.00cm, -0.5cm)$) ;

\draw[<-](Wind.west) -- +(-0.5cm, 0.0cm)node[above, name = n]{$d$};



\end{tikzpicture}

	
        \caption{$Q\alpha$ analysis interconnection}
        \label{Fig:AnalysisConn}
\end{figure}

In Fig. \ref{Fig:AnalysisConn}, the interconnection used for the LTV analysis is shown. 
The ELV's dynamics are described by the respective LTV system $G_t$. All other blocks correspond to the systems discussed in the preceding subsections.
On the contrary, the Monte Carlo simulation directly utilizes the nonlinear launcher's dynamics as given in (\ref{eq:PitchNonLin}), and the wind model for the nonlinear analysis is applied. Both analyses apply the same nominal TVC model $G_\text{TVC}$ and PID controller $C$.

\

\subsection{Analysis Results}\label{ss:Results}
The LTV worst-case aerodynamic load $Q\alpha_\text{WC}$ is calculated, solving the corresponding optimization problem (\ref{OptiProblem}) using Algorithm 1. Due to the definition of the $L_2[0,T]$-to-Euclidean gain, the value of $Q\alpha$ is only upper bounded at the final time of the analysis horizon. Therefore, an analysis over a set of final times $T_i$ covering the trajectory is necessary to identify $Q\alpha_\text{WC}$. An evenly spaced set of final times $T_i$ spanning from $30\text{s}$ to $95s$ with a step size of $5\text{s}$ is chosen. 

The interconnection in Fig. \ref{Fig:AnalysisConn} must be transformed into the IQC framework as specified in Section \ref{ss:LTVRob}. The individual uncertainties given in Table \ref{tab:UncSet} are replaced by the IQC description given in Example \ref{ex:delta} and \ref{ex:Delta}, respectively, and stacked diagonally in a single IQC. The respective values for the norm bound $b$, as well as the parameter of the basis function $\nu$ and $\rho$, are given in Table \ref{tab:IQCparam}.
\begin{table}\label{tab:IQCSettings}
\caption{Parameters used for the IQC \label{tab:IQCparam}}
\centering
\begin{tabular}
{ p{1.7cm} p{1.7cm} p{1.7cm} p{1.7cm}}
 Uncertainty & $b$ &$\nu$ & $\rho$ \\
 \hline
 $\Delta_\text{TVC}$	&$1$		& $1$ 		&$-1$ \\
 $\delta_{l_{GA}}$			&$0.2$		& $1$		&$-1$\\
 $\delta_{l_{CG}}$ 	&$0.1$  		& $1$ 	 	&$-1$\\
 \hline
\end{tabular}
\end{table}

For calculating $Q\alpha_1$ for $T_1 = 30\text{s}$,  Algorithm 1 is run with an initial scaling of the uncertainty norm bounds by $k_\text{IQC}=0.3$.
The lower bound $m$ and upper bound $n$, i.e. the exponent range, for each logarithmic search variable $\bar{\zeta}_i$ is set to $-4$ and $9$, respectively. This corresponds to a search space from $-10^9$ to $10^9$ with the smallest absolute value of $10^{-4}$.
As $X>0$, the diagonal entries of $X$ must be strictly positive. Thus,  respective elements of $\bar{\xi}$ in the solution vectors are fixed to $0$. The single IQC parameterizations combine for a total of $23$ decision variables.

The value for the maximal population size $n_\text{P,max}$ is set to $50$.
$n_\text{P,min}$ is set to $4$. The maximum number of iterations is set to $i_\text{max}=30$.
Both the number of successful scaling factors $k_{F}$ and the number of successful crossover rates $k_{CR}$ are set to $5$.
The value for $\gamma_\text{lim}$ is set to the nominal value $\gamma_\text{nom}$ of the respective final time.
Here, $\gamma_\text{nom}$ is the nominal worst-case gain of the interconnection for the final time $T_1 = 30\text{s}$. It is calculated by the nominal formulation of Theorem 1, using the procedure proposed by \cite{Green1995}. Note that $\gamma_\text{lim}$ can be chosen to a higher value if only the compliance to said value needs to be shown.
The bisection's lower bound is set to $\gamma_\text{LB}=\gamma_\text{nom}$ and the upper bound to $\gamma_\text{UB}=100\gamma_\text{nom}$. Furthermore, a relative tolerance of the bisection of $\epsilon_\text{BS}=10^{-5}$.

For following terminal times $T_i$, Algorithm 1 is called adding the optimal solution of the previous terminal time $T_{i-1}$ to the initial population. 
Additionally, the search space is also reduced to $\pm2$ the magnitude of the values of the previous best solution.
A downscaling of the IQC norm bounds is no longer necessary, i.e. $k_\text{IQC}=1$. These adjustments exploit that the optimal solution of consecutive final times $T_i$ and $T_{i+1}$ are relatively close. Consequently, the overall computational effort can be reduced for all following times. This marks one of the major advantages over the LMI/SDP-based approaches.
The remaining inputs of Algorithm 1 are set to $n_\text{P,max}=30$, $n_\text{P,min}=4$, $i_\text{max}=20$, $k_{CR} = 5$, and $k_{F}=5$.

The whole interval can be solved using the proposed algorithm on average in $2\text{h} 55\text{min}$ over seven runs on a standard computer with an Intel Core i7 processor and $32$GB memory. For the analysis, the bisection step of the optimization was parallelized on eight physical cores.
The maximum worst case load $Q\alpha_\text{WC,max}$ calculated for the trajectory is $29.9\%$ of the limit load $Q\alpha_\text{lim}=220000\text{Pa}^\circ$
occurring at $30\text{s}$ after lift-off. It correlates with the highest expected turbulence intensities.

The turbulence is inversely altitude-dependent, reaching its highest values between $30\text{s}$ and $45\text{s}$ after lift-off, corresponding to altitudes of $4.66\text{km}$ and $9.95\text{km}$, respectively. Simultaneously, the dynamic pressure build-ups from $4.52\cdot 10^4Pa$ to $5.47\cdot10^4\text{Pa}$, resulting in the flight's highest aerodynamic loads. However, when the dynamic pressure reached its maximum  $Q_\text{max}=5.603\cdot10^4\text{Pa}$ at $52\text{s}$ after lift-off ($12.9\text{km}$), the turbulence intensity reduced to $\sigma_u = 0.05\text{m/s}$, and the wind turbulence is negligible.

A comparison with the corresponding nominal LTV worst case-analysis shows only a marginal impact of the uncertainties. The nominal worst-case load of $Q\alpha_\text{nom,max} = 28.27\%$ occurring at $30\text{s}$ after lift-off is just insignificantly less than $Q\alpha_\text{WC,max}$. Thus, it can be concluded that the external disturbance rather than the perturbation in the launcher's parameters is the key influence on the expected worst-case loads.
The resulting maximum loads of the nominal and worst-case LTV nominal analyses are summarized in the first row of Tab. \ref{tab:QalphaComp}.

\begin{table}\label{tab:QalphaComp}
\caption{Comparison of maximal values of $Q\alpha$ due to wind turbulence \label{tab:QalphaComp}}
\centering
\begin{tabular}
{p{1.2cm} p{2.0cm} p{2.0cm} p{2.0cm}}
 Analysis 	& Nominal 				& Fixed Perturbation  		& Worst Case \\
 \hline
 LTV 	&$28.27\%$	 	& 	$29.2\%$ 	& $29.9\%$\\
 Monte Carlo 		&$19.6\%$	 	& 	$22.0\%$ 	&  - \\
 \hline
\end{tabular}
\end{table}
A Monte Carlo simulation is run on the nonlinear launcher model to validate the worst-case analysis results. It directly applies the Dryden filter (\ref{eq:DrydenNonLin}). Instead of a dynamic TVC uncertainty, the corresponding parametric uncertainties specified in Section  \ref{ss:UncMod} are directly implemented.
One thousand white noise input disturbances are considered, generated by the Simulink internal band-limited white noise block with uniformly distributed random noise seeds. The parametric uncertainties are uniformly gridded over their definition space, i.e.  $5$ points to cover $\pm20\%$ uncertainty in $l_{GA}$ and $5$ points to cover $\pm 10\%$ uncertainty $l_{CG}$.
Finally, the static gain, eigenfrequency, and damping ratio of the TVC are gridded over $\pm 10\%$ with three equidistant points. Additionally, two different time delays are considered, namely 0.005s and 0.01s. 
Subsequently, each of the resulting $1350$ models is evaluated for every noise signal $n_{w_i}(t)$.  A single execution of the nonlinear simulation takes an average of $1.5\text{s}$, resulting in an overall analysis time of ca. $15$ and a half days for a relatively coarse analysis grid. 
It is reduced to an effective time of $3\text{d}22\text{h}$, distributing the analysis to four computers equipped with Intel Xeon E-5 1620 v4 processors and  $32$GB memory.

\begin{figure}[ht!]
\centering
\begin{tikzpicture}
\definecolor{blue1}{RGB}{222,235,247}
\definecolor{blue2}{RGB}{158,202,225}
\definecolor{blue3}{RGB}{49,130,189}
\begin{axis}[ width = 1.0\columnwidth, height = 0.6\columnwidth, 
  	grid=major, 
   grid style={densely dotted,white!60!black}, 
   xlabel= Time $t  \,{[s]}$, 		
   ylabel= Scaled $Q\alpha\,{[-]}$, 	
   legend style={at={(0.65,0.97)},anchor=north west},
   legend cell align = {left},
   xmin = 30, xmax = 95, ymin = -0.5, ymax = 0.5, 
        ]

\addplot[red!60, line width = 1.5,  no marks] table[x expr = \thisrowno{0} ,y expr = \thisrowno{9} ,col sep=comma] {figures/E2PPlotFinal_final_new.csv};\label{LTVBoundUpp}
\addplot[red!60, line width = 1.5,  no marks] table[x expr = \thisrowno{1} ,y expr = \thisrowno{10} ,col sep=comma] {figures/E2PPlotFinal_final_new.csv};\label{LTVBoundUpp}

\addplot[blue1, line width = 0.75,  no marks] table[x expr = \thisrowno{4} ,y expr = \thisrowno{13} ,col sep=comma] {figures/E2PPlotFinal_final_new.csv};\label{QalphaSignalAll}

\addplot[blue1, line width = 0.75,  no marks] table[x expr = \thisrowno{6} ,y expr = \thisrowno{15} ,col sep=comma] {figures/E2PPlotFinal_final_new.csv};

\addplot[blue1, line width = 0.75,  no marks] table[x expr = \thisrowno{6} ,y expr = \thisrowno{15} ,col sep=comma] {figures/E2PPlotFinal_final_new.csv};

\addplot[blue1, line width = 1,  no marks] table[x expr = \thisrowno{7} ,y expr = \thisrowno{17} ,col sep=comma] {figures/E2PPlotFinal_final.csv};

\addplot[black, line width = 0.75,  no marks] table[x expr = \thisrowno{5} ,y expr = \thisrowno{14} ,col sep=comma] {figures/E2PPlotFinal_final_new.csv};\label{QalphaSignalAllWC}

\addplot[red!60, dashed,line width = 1.5,  no marks] table[x expr = \thisrowno{2} ,y expr = \thisrowno{11} ,col sep=comma] {figures/E2PPlotFinal_final_new.csv};\label{MCBoundUpp}
\addplot[red!60, dashed,line width = 1.5,  no marks] table[x expr = \thisrowno{3} ,y expr = \thisrowno{12} ,col sep=comma] {figures/E2PPlotFinal_final_new.csv};

\end{axis}
\end{tikzpicture}
\caption{Analysis results scaled with $Q\alpha_\text{lim}$: $Q\alpha_\text{WC}$ bound LTV analysis (\ref{LTVBoundUpp}),  $Q\alpha_\text{MC}$ bound Monte Carlo simulation (\ref{MCBoundUpp}),  $Q\alpha_\text{MC}$ signals (\ref{QalphaSignalAll}),
$Q\alpha_\text{MC,max}$ signal Monte Carlo Simulation  (\ref{QalphaSignalAllWC})}
\label{fig:results_E2P}
\end{figure}
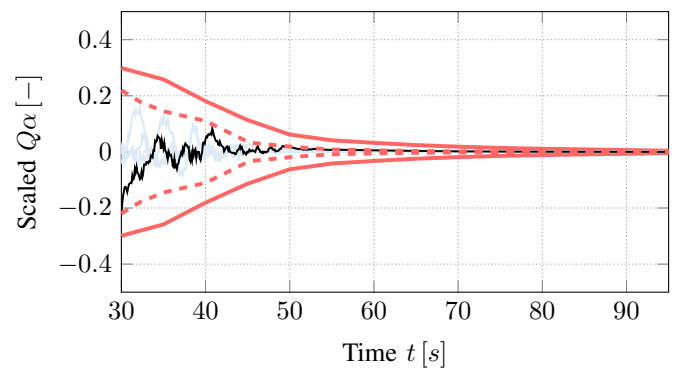

The nonlinear simulation starts at $t_s = 25\text{s}$ and ends at $t_f = 95\text{s}$ after lift-off. 
A maximum aerodynamic load $Q\alpha_\text{MC,max}$ of $22.0\%$ of $Q\alpha_\text{lim}$ is identified at $32.1\text{s}$, with the corresponding uncertainty description as follows:
$\delta_{l_{CG}} = -0.1$, $\delta_{l_{GA}}=0.2$, $\delta_k= -0.1$, $\delta_{\omega}=-0.1$, $\delta_\zeta=-0.1$, and $\tau=0.01\text{s}$.
The respective $Q\alpha_\text{MC,max}$ signal scaled by the limit load $Q\alpha_\text{lim}$ is shown in Fig. \ref{fig:results_E2P}, where it is compared to the LTV worst-case aerodynamic loads. Note the latter's values in-between the analysis grid points $T_i$ are linearly interpolated.
Furthermore, an envelope covering the peaks of all simulated $Q\alpha_\text{MC}$ signals and the corresponding signals are plotted in Fig. \ref{fig:results_E2P}. The former begins at $T_1=30\text{s}$, corresponding to the LTV analysis, by which it is upper bounded for the whole trajectory. 

As for the LTV case, a nominal simulation is conducted to assess the influence of the uncertainties. Therefore, nominal nonlinear ELV is evaluated for all 
$n_{w_i}(t)$. It provides a maximum nominal load $Q\alpha_\text{MC,nom}$ of $19.6\%$ of the limit load, showing little influence of the perturbations. The maximum $Q\alpha$ values of both Monte Carlo simulations are summarized in Tab. \ref {tab:QalphaComp}.

For the gap between the LTV and the nonlinear analysis, the fundamental difference in both analyses' nature is the leading cause. The worst-case LTV analysis calculates a guaranteed upper bound for $Q\alpha$ of the interconnection in Fig. \ref{Fig:AnalysisConn}. On the contrary, the Monte Carlo analysis can only provide a lower bound.
Furthermore, in the LTV analysis, the disturbance input is a worst-case norm bounded signal, whereas, in the nonlinear simulation, it is an arbitrary band-limited white noise signal. The resulting wind disturbance signals have a comparable PSD due to the LTV wind filter design in Section \ref{sec:Example}. However, it is unlikely that the exact LTV worst-case wind signal will be under the evaluated signals in the Monte Carlo simulation. 
Another contributor is dynamic uncertainty $\Delta$, which likely introduces some conservatism in the LTV analysis compared to the nonlinear analysis's parametric uncertainties.

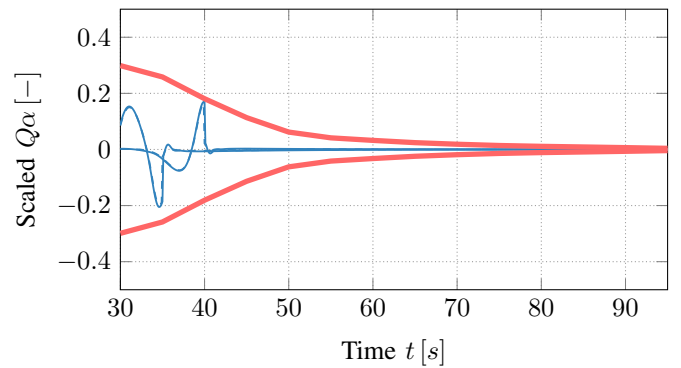
\begin{figure}[ht!]
\centering
\begin{tikzpicture}
\definecolor{blue1}{RGB}{222,235,247}
\definecolor{blue2}{RGB}{158,202,225}
\definecolor{blue3}{RGB}{49,130,189}
\begin{axis}[ width = 1.0\columnwidth, height = 0.6\columnwidth, 
  	grid=major, 
   grid style={densely dotted,white!60!black}, 
   xlabel= Time $t  \,{[s]}$, 		
   ylabel= Scaled $Q\alpha\,{[-]}$, 	
   legend style={at={(0.65,0.97)},anchor=north west},
   legend cell align = {left},
   xmin = 30, xmax = 95, ymin = -0.5, ymax = 0.5, 
        ]

\addplot[blue3, line width = 0.7,  no marks] table[x expr = \thisrowno{2} ,y expr = \thisrowno{11} ,col sep=comma] {figures/E2PdWCplot_new_new.csv};\label{QalphaLTVdWC}
\addplot[blue3, dashed,line width = 0.7,  no marks] table[x expr = \thisrowno{3} ,y expr = \thisrowno{12} ,col sep=comma] {figures/E2PdWCplot_new_new.csv};\label{QalphaNLdWC}

\addplot[blue3, line width = 0.7,  no marks] table[x expr = \thisrowno{4} ,y expr = \thisrowno{13} ,col sep=comma] {figures/E2PdWCplot_new_new.csv};
\addplot[blue3, dashed,line width = 0.7,  no marks] table[x expr = \thisrowno{5} ,y expr = \thisrowno{14} ,col sep=comma] {figures/E2PdWCplot_new_new.csv};

\addplot[red!60, line width = 2,  no marks] table[x expr = \thisrowno{0} ,y expr = \thisrowno{9} ,col sep=comma] {figures/E2PdWCplot_new_new.csv};\label{LTVBound}
\addplot[red!60, line width = 2,  no marks] table[x expr = \thisrowno{1} ,y expr = \thisrowno{10} ,col sep=comma] {figures/E2PdWCplot_new_new.csv};

\end{axis}
\end{tikzpicture}
\caption{Analysis results scaled with $Q\alpha_\text{lim}$: $Q\alpha_\text{WC}$ bound LTV IQC analysis (\ref{LTVBound}), $Q\alpha(t)$ LTV simulation (\ref{QalphaLTVdWC}),
$Q\alpha(t)$ for $d_\text{WC}(t)$ nonlinear simulation (\ref{QalphaNLdWC})  }
\label{fig:dWC}
\end{figure}

Concluding, the compliance of the LTV and nonlinear launcher model is evaluated.
Therefore, the worst-case disturbance signals of the LTV analysis interconnection in Fig. \label{Fig:UncIntLTV} for a fixed uncertainty combination are calculated using the approach proposed in \cite{Iannelli2019} for every $T_i$. 
The uncertainty combination corresponds to $Q\alpha_\text{MC,max}$, namely $\delta_{l_{CG}} = -0.1$, $\delta_{l_{GA}}=0.2$, $\delta_k= -0.1$, $\delta_{\omega}=-0.1$, $\delta_\zeta=-0.1$, and $\tau=0.01\text{s}$. A second-order Pad\'e approximation is used to approximate the behavior of the time delay.
Afterwards, the LTV analysis interconnection is simulated with the calculated worst-case disturbance signals $d_{\text{WC},i}(t)$. For a better application in the nonlinear model, each $d_{\text{WC},i}(t)$ is evaluated along the complete trajectory starting from $25s$ and ending at $95s$ with
\begin{equation}
\label{eq:d_WC}
d_{\text{WC},i}(t) := 
\begin{cases}
   d_{\text{WC},i}(t) & \text{for } t \le T_i\\    
  0    & \text{for } t > T_i
\end{cases}
.
\end{equation}
Afterwards, the recorded LTV worst-case wind signals $w_{\text{WC},i}$ are used as input in the simulation of the accordingly perturbed nonlinear model.
In Fig. \ref{fig:dWC}, the resulting $Q\alpha(t)$ signals for $T=35\text{s}$ and $T=40\text{s}$ of the LTV and nonlinear simulation are compared.
An evaluation against the worst-case envelope $Q\alpha_\text{WC}$ shows no violation. Furthermore, the LTV and nonlinear simulation results match closely. Together with the narrowed gap to the worst-case LTV analysis, the general accuracy of the LTV analysis framework for the given uncertainties is validated. 

\section{Conclusion}
The presented analysis framework offers an efficient approach to calculate the worst-case gain of uncertain finite horizon LTV systems. A novel algorithm efficiently exploits the structure of the optimization problem facilitating the analysis of industry-sized problems. 
Its applicability is demonstrated using an LTV worst-case loads analysis of a space launcher, validated against a Monte Carlo simulation conducted on the corresponding nonlinear model.
The LTV analysis provides a valid, not overly conservative upper bound on the Monte Carlo simulation of the nonlinear model in a fraction of time.
Hence, the presented analysis framework provides a valuable supplemental tool for certification processes.

\section*{Acknowledgment}
The authors would like to thank the ESA for the financial support through the Networking/Partnering Initiative with contract number 4000123233.

\ifCLASSOPTIONcaptionsoff
  \newpage
\fi



%

\bibliographystyle{IEEEtran}
\bibliography{Ref_LauncherIQCAnalysis}



%

\begin{IEEEbiography}[{\includegraphics[width=1in,height=1.25in,clip,keepaspectratio]{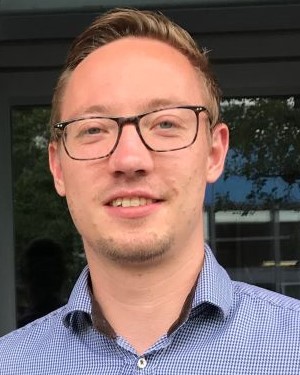}}]{Felix Biert\"umpfel}
Felix Biert\"umpfel received his M.Sc. degree from Technical University of Munich, Germany, in 2016. From 2017 to 2021, he was a Ph.D. student at University of Nottingham. Since 2021 he works as a research associate at the Technische Universit\"at Dresden at the Chair of Flight Dynamics and Control. His research interest is mainly in robust and linear time varying-control for space applications.
\end{IEEEbiography}

\begin{IEEEbiography}[{\includegraphics[width=1in,height=1.25in,clip,keepaspectratio]{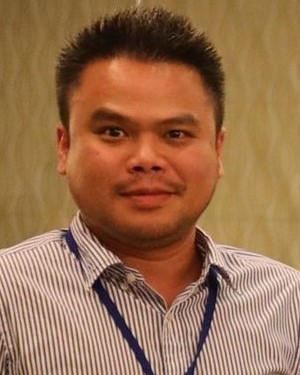}}]{Nantiwat Pholdee}
Nantiwat Pholdee received the B.Eng and Ph.D. in mechanical engineering from Khon Kaen Univeristy, Thailand, in 2008 and 2013, respectively. Currently, he is an associate professor at the department of mechanical engineering, Khon Kaen Univeristy. His research interests include multidisciplinary design optimization, meta-heruistic algorithms and flight dynamics and control.
\end{IEEEbiography}

\begin{IEEEbiography}[{\includegraphics[width=1in,height=1.25in,clip,keepaspectratio]{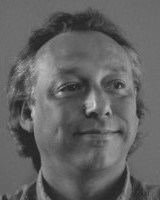}}]{Samir Bennani}
Samir Bennani received the Ph.D. degree from the Aerospace Engineering Department, Delft University of Technology, Delft, The Netherlands. Since 2005, he serves as a senior advisor in the Control System Division at European Space Agency, Noordwijk, The Netherlands. His main research activities are concentrated around the industrialization of space systems of advanced and robust control techniques. Dr. Bennani is a Senior Member of AIAA and is part of the various Technical Committees.
\end{IEEEbiography}

\begin{IEEEbiography}[{\includegraphics[width=1in,height=1.25in,clip,keepaspectratio]{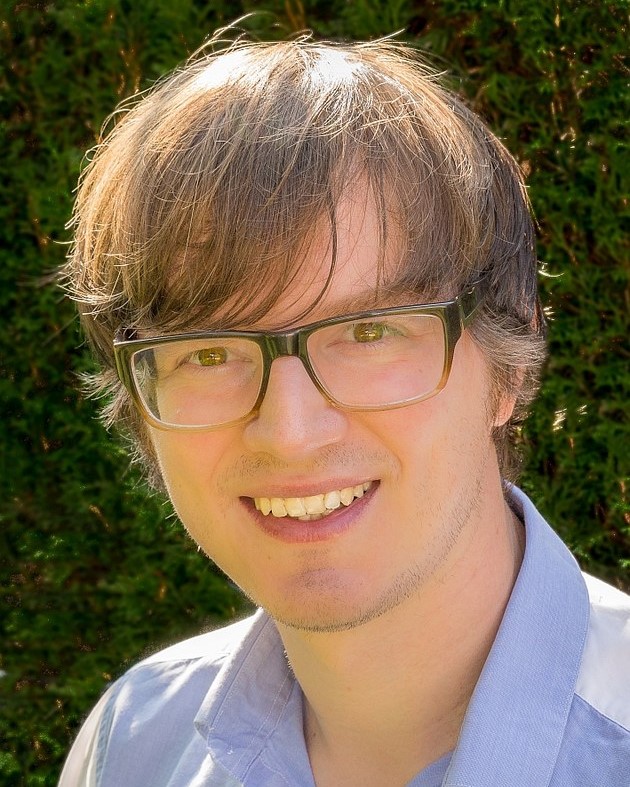}}]{Harald Pfifer}
Harald Pfifer is the Chair in Flight Dynamics and Control at the Technische Universit\''at Dresden, Germany. He received his Ph.D. from the Technical University Munich, Germany, in 2013. Before joining the Technische Universit\''at Dresden in 2021, he had been an assistant professor at University of Nottingham, and a post-doctoral associate at University of Minnesota. His main research interests include aero-servo-elastic control, modeling of uncertain dynamical systems, and robust and linear parameter-varying control.
\end{IEEEbiography}





\end{document}